# The dynamic chirality flips of Skyrmion bubbles

Yuan Yao[1,4]*, Bei Ding[1,4], Jinjing Liang[1,3], Hang Li[1], Xi Shen[1], Richeng Yu[1], Wenhong Wang[1,2,3*],

[1] Beijing National Laboratory for Condensed Matter Physics, Institute of Physics, Chinese Academy of Sciences, Beijing 100190, China

[2] Songshan Lake Materials Laboratory, Dongguan, Guangdong 523808, China

[3] University of Chinese Academy of Sciences, Beijing 100049, China

[4] These authors contributed equally: Yuan Yao and Bei Ding

**Corresponding Author**
*E-mail: yaoyuan@iphy.ac.cn
  wenhong.wang@iphy.ac.cn

**Abstract**

   Magnetic skyrmion, a topological magnetic domain with complex non-coplanar spin texture, appears a disk-like structure in two dimensions. Exploring three-dimensional spin texture and related chirality switching has drawn enormous interests from the perspective of fundamental research. Here, the three-dimensional magnetic moment of the skyrmion bubbles in centrosymmetric Mn-Ni-Ga were reconstructed with the vector field tomography approach via Lorentz transmission electron microscopy. The type of the bubbles was determined from investigating the magnetic vectors in entire space. We found that the bubbles switched their chirality easily but still keep the polarity to remain the singularity of the bubbles within the material. Our results offer valuable insights into the fundamental mechanisms underlying the spin chirality flips dynamics of skyrmion bubbles.

**Introduction**

Owing to small size and fast mobility, skyrmion, a novel topological magnetic structure becomes an attractive research topic recently for potential application in future memory device[1-4]. Generally, skyrmions are stabilized in noncentrosymmetric systems with a uniform chirality because of Dzyaloshinskii-Moriya interaction (DMI)[2,5,6] but can also be formed in centrosymmetric magnet under competition between dipole-dipole interaction (DDI) and magnetic anisotropy[7-9]. The DDI skyrmions are the traditional named magnetic bubbles, or the fashionable skyrmion bubbles. Either DMI or DDI skyrmion contains two degrees of freedom, polarity ($p$) and vorticity ($w$), determining the chirality of the skyrmion. Considering the stability, DMI skyrmion is regarded as a favorite candidate for the spintronic application although the operation of DMI skyrmion need to reverse the whole swirl accompanied with its polarity owing to the intrinsic DMI[10]. Contrary to DMI-stabilized skyrmions that exhibit the collective chirality and polarity motivated by their intrinsic DMI, skyrmion bubbles in centrosymmetric magnets always demonstrate the individual feature and deform themselves randomly because DDI could not strongly couple the bubbles to coordinate their local magnetic behavior. The most interesting physics in skyrmion bubbles is that the vorticity of their spin textures varies with the internal structure of Bloch lines (BLs), resulting in a variety of spin textures.

Lorentz transmission electron microscopy (LTEM) turns into a powerful tool to characterize skyrmions and has disclosed diverse topological configurations with its high spatial resolution. By virtue of the *in situ* techniques, for instance, cooling, heating or extra electron-magnetic field, the transformation and the movement of the skyrmions can be observed directly in LTEM[11-13]. However, even without the artifacts from the specimen orientation or image processing[14,15], the magnetic moments retrieved from transport of intensity equation (TIE)[6], holography[16,17] and differential phase contrast (DPC) methods[18,19] still conceal some real features since the LTEM images are a projection of the magnetic structures. Figure 1 depicts this shortage to observe a simulated Bloch skyrmion bubbles with different spin configurations. Two parameters – $p$ and $w$- were used to describe the skyrmion bubbles. In a given coordinate system, the spins at cores of the bubbles can point either up (*p*+) or down (*p*-), while the in-plane magnetization can rotate clockwise (*w*+) or anticlockwise (*w*-). The polarity of skyrmion bubbles can be further identified by the surface twisting, where *p*+ for a divergent Néel twisting at top side and a convergent Néel twisting at bottom side and *p*- with an opposite situation. Both polarity and vorticity determine the charity of skyrmion bubble, L+ (*p*+, *w*+), L- (*p*-, *w*-), R+ (*p*+, *w*-) or R- (*p*-, *w*+), as illuminated in Figure 1a. Clearly, clockwise skyrmion bubbles (*w*+) regardless of their polarity display the same magnetic contrast in LTEM images and then demonstrate identical direction mapping of in-plane moments. The similar phenomena appear in anticlockwise skyrmion bubbles with *p*+ and *p*-. Obviously, LTEM images or processing methods alone fails to discern the real spin configuration of the skyrmions or to clarify the variation of spin texture when the bubble contrast varies. Thus, the characterization of three-dimensional (3D) magnetic structure is a strong demand for thoroughly understanding the static feature or dynamic process of skyrmions in the magnets.

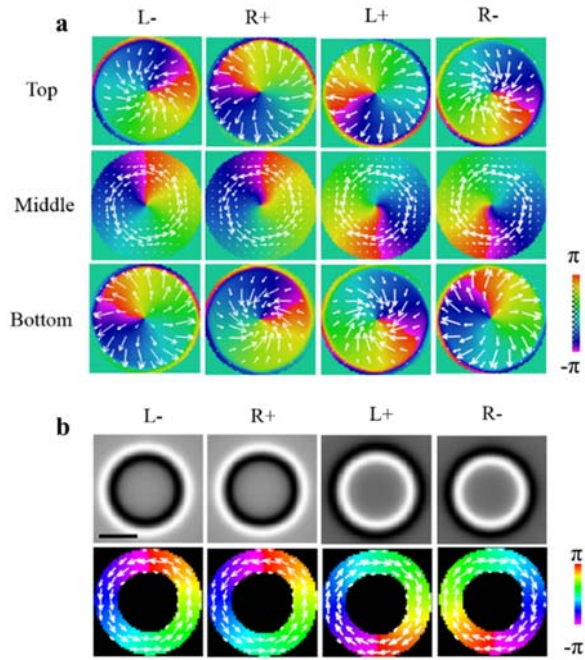

**Fig.1. Simulation of the skyrmion bubbles. a** Characteristics of different type skyrmion bubbles and their moment feature at surface and middle of the TEM specimen. **b** The corresponding simulated over-focus Lorentz-TEM images and the recovered in-plane

magnetization distribution map from the simulated Lorentz-TEM images (color denotes the direction of the local spin). Scale bar is 50 nm.

Recently, several techniques have been employed to disclose the detailed 3D spin texture of a DMI skyrmion in noncentrosymmetric systems, such as resonant elastic x-ray scattering (REXS)[20,21] and associated scanning X-ray magnetic circular dichroism (XMCD)[22]. In real space, tomography in TEM is a proper approach to visualize the 3D morphology of the specimen and yields plentiful fruits in materials characterization[23]. However, different from the reconstruction of morphology or chemical concentration features where the contribution of each voxel to the image contrast does not change during specimen rotation, the 3D recovery of magnetic structure is more complicated because it is a vector field where each voxel can vary with its projected intensity in image at different orientation, depending on the dot product between the vector and the projection line[24,25]. A smart solution is as the component of a vector along the titling axis always provides a constant contribution to the image contrast (dot production does not change when tilting specimen), the 3D feature of this component can be retrieved correctly from a series of tilting images. Then three orthogonal components, $B_x$, $B_y$ and $B_z$ are reconstructed separately by tilting the specimen around corresponding axes and entire vector field may be recovered with these components. Fortunately, owing to the physical fact $\nabla \cdot \vec{B} = 0$, only two orthogonal tilting series, such as $B_x$ and $B_y$, are enough to rebuild the entire 3D magnetic vector field with the proper boundary conditions[26-29]. Wolf et al brilliantly described the entire roadmap in great detail[29]. With this algorithm some micro-magnetic features have been successfully portrayed, including the DMI skyrmions[30].

In this work, we acquired multi-angle LTEM image series around x- and y-axis to reveal the internal 3D magnetic configuration of magnetic bubbles within centrosymmetric MnNiGa via tomography technique. Integrating reconstructions and micro-magnetic simulations uncovered real-space 3D spin configuration of different type of bubbles. Without information of the out-plane component and by analyzing the surface Néel twisting of bubbles, we can determine the chirality of the skyrmion bubbles and found that these bubbles can change their in-plane magnetic moment easily while remain their polarity. The readily reversal of the chirality without polarity flipping facilitates the application of DDI skyrmions as the feasible memory device.

**Results**

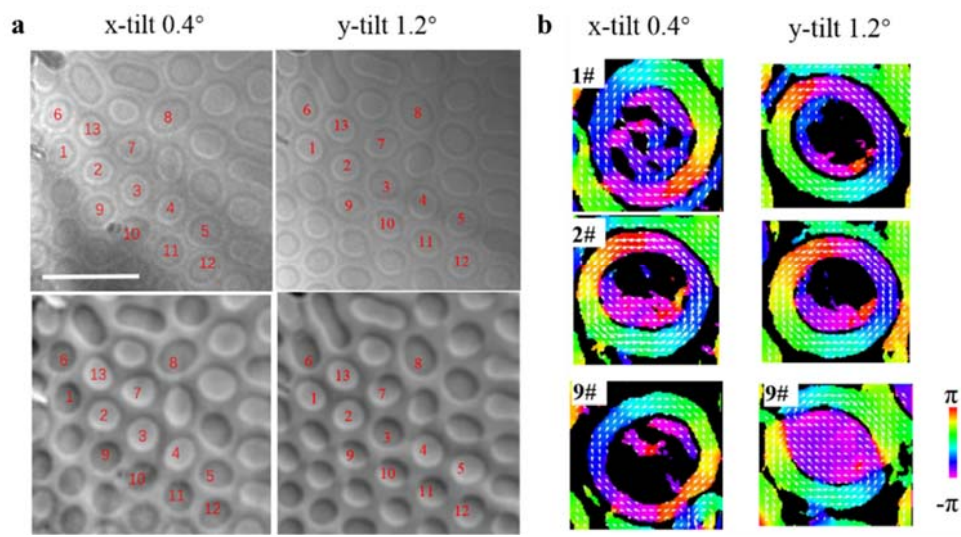

**Fig. 2 Real-space observation of field free room temperature skyrmion bubbles after FC manipulation. a** The 300 μm over-focus images (first row) and TIE retrieved phase image (second row) of x-tilt 0.4° and y-tilt 1.2°, respectively. Scale bar is 500 nm. **b** The configuraiton of magnetic moments of 1#, 2# and 9# bubbles, deduced from the phase images in a).

The tomography reconstruction of magnetic spin configuration was performed on the centrosymmetric magnet $(Mn_{1-x}Ni_x)_{65}Ga_{35}$ (X=0.45), which was the same as that used in our previous work[31]. The reason for choosing MnNiGa is that the field free magnetic bubbles can be stabilized at room temperature[11,12] and related detailed experimental process are shown in Methods. Fig. 2a shows the over-focus images and retrieved phase maps of field free skyrmion bubbles at $\theta_x = 0.4°$ and $\theta_y = 1.2°$, respectively. When $\theta_x = 0.4°$, the occlusive magnetic contrast represents apparently the typical skyrmion bubbles. Relatively, some bubbles in y-tilt image changed their contrast or shape, which also are manifested by the black-white reversals in retrieved phase images definitely. The magnetic moment directions deduced from the phase image directly reveal the in-plane spin configuration of each bubble and imply the variation of bubbles, such as 1#,

2# and 9# shown in Fig. 2b. However, owing to the absence of out-plane spin configuration, it's hard to verify the real chirality of each bubble from the direction mapping in Figure 2b since the variation of vorticity or polarity can both result in the observed phenomena, as illustrated in Figure 1, and it is impossible to clarify further what motivated such variation. For example, if the vorticity and polarity varied simultaneously, the final state also appeared as the initial feature as of 2# bubble. Furthermore, compared to the x-tilt image the shapes of most bubbles in y-tilt image distorted severely, it also rules out the possibility to compute $B_z$ component of these bubbles because the voxel in $B_x$ matrix cannot be registered correctly to the $B_y$ matrix so that equation $\nabla \cdot \vec{B} = 0$ is not available.

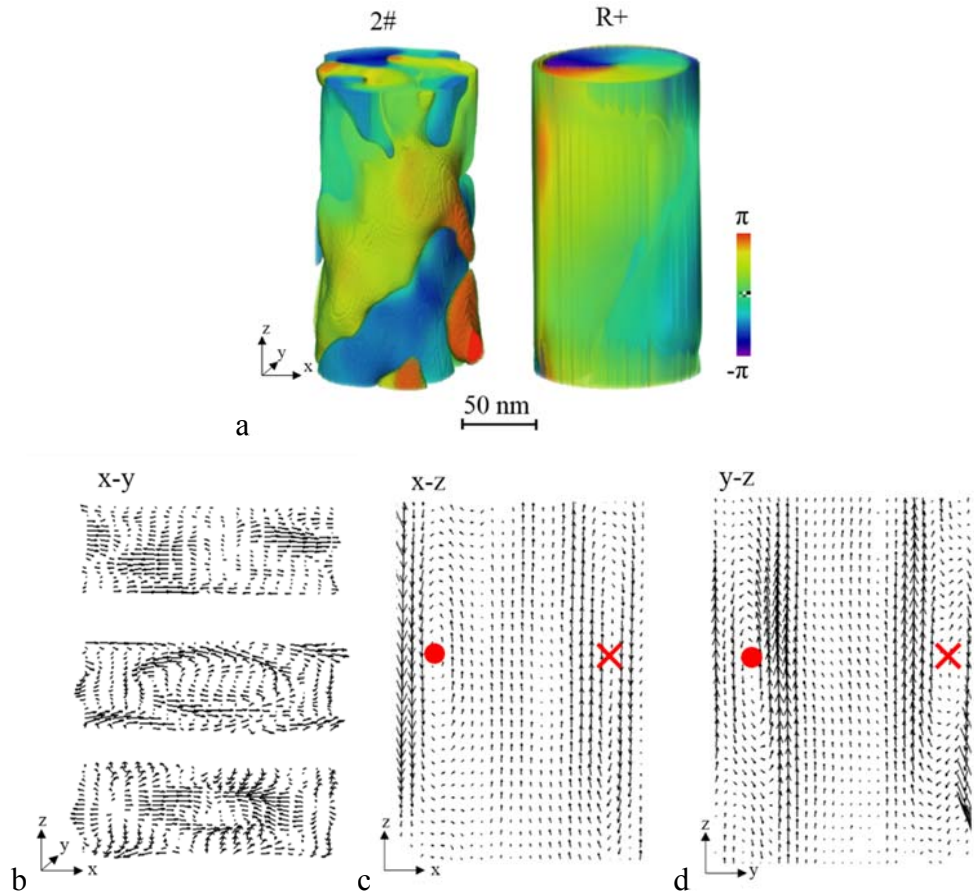

**Fig.3 The magnetic R+ configuration of 2# bubble. a** Orientation distribution of in-plane magnetic moments in 3D (color rending) feature (left) with the R+ model (right), **b** moment components within the top, middle and bottom x-y section; **c, d** moment components lying on middle x-z section and y-z section. The Bloch lines revolving around the waist of bubble are denoted in c) and d).

In the following, we analyze comprehensively those sets of 3D magnetic induction data in order to extract characteristic magnetic features and quality of bubbles in MnNiGa. First, a little shape-changed bubble 2# in Figure 2 is optimal to testify the feasibility of vector field reconstruction, the detailed analysis are shown in Methods. Figure 3a interprets the rendered moment orientation of in-plane component in three dimensions with the referred R+ model and Figure 3b-d outline the corresponding moments in XY, YZ and XZ plane, respectively. Obviously, the 3D texture of 2# bubble successfully replicates a R+ skyrmion bubble, which not only justifies the reliability of the reconstruction approach for the micro-magnetic object but also confirms that bubbles in the real lamina contain the surface states in the boundaries, as predicted by the theory[32]. During the tomography data acquisition, it is difficult to align the out-plane component of the bubbles to the electron beam precisely. Leaning towards x-axis or y-axis of the laboratory frame could lead that the out-plane component contributes to the notable $B_x$ or $B_y$ components in the center of the retrieved bubbles displayed in Figure 2b and Figure 3c.

A more detail insight into the simulation and experimental data gives the hints that the polarity of bubbles can be inferred from the reconstructed surface features i.e., the convergent or divergent manner, without calculating $B_z$ component, to determine the type of bubbles (Figure 1a and 3b). Furthermore, we also notice that only one set of data exhibiting radial or anti-radial nature on the surface provides enough evidence to deduce the polarity of the bubbles, based on the prior knowledge of the bubbles. Thus, the type of bubbles in Figure 1 can be registered by their independent 3D matrix of $B_x$ or $B_y$ component. In deed it should be a more convinced method to combine the swirling character revealed by TIE approach in fig. 2b, which further moderates the demands on data acquisition for discriminating the bubbles.

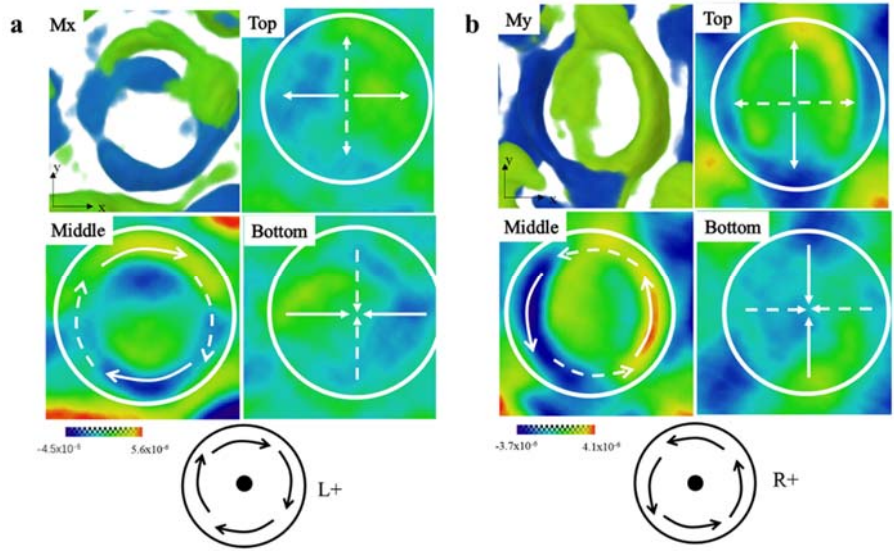

**Fig. 4 The magnetic spin transition of 5# bubble.** The feature of x component (**a**) and y component (**b**) of magnetic moments at top, middle and bottom x-y sections, respectively. Solid arrows portray the directions for the corresponding component while dashed arrows are the speculative direction of orthogonal component based on the prior knowledge of Bloch skyrmion bubbles. Color legend indicates the relative magnitude of magnetic moment.

Base on the experimental verification, the corresponding inference was applied to distinguish the different type of bubbles which changed their magnetic texture at two series where only $B_x$ or $B_y$ component were reliably reconstructed independently. The in-plane $B_x$ components of 5# bubble in x-tilt experiment is convergent near the bottom, clockwise in the middle part and divergent beneath the top surface. According to the models in Figure 1, it should be L+ (*p+, w+*) state. Whereas in y-tilt situation, the in-plane $B_y$ components are convergent, counter clockwise and divergent, belong to the R+ (*p+, w-*) type. Undoubtedly, 5# bubble has reversed its spiral direction but held polarity unchanged. Although the detail of the magnetic moment configurations within the bubble is not clear, the magnetic structure transformation between x-tilt and y-tilt could be tackled correctly.

|     | X-tilting | | | | | | Y-tilting | | | | | |
| --- | --- | --- | --- | --- | --- | --- | --- | --- | --- | --- | --- | --- |
|     | Top | Middle | Bottom | p | w | Type | Top | Middle | Bottom | p | w | Type |
| 1#  | D | CCW | C | + | - | R+ | D | CW | C | + | - | L+ |
| 2#  | D | CCW | C | + | - | R+ | D | CCW | C | + | + | R+ |
| 3#  | D | CCW | C | + | - | R+ | D | CW | C | + | - | L+ |
| 4#  | D | CCW | C | + | - | R+ | D | CCW | C | + | + | R+ |
| 5#  | D | CW | C | + | + | L+ | D | CCW | C | + | + | R+ |
| 7#  | D | CCW | C | + | - | R+ | D |    | C | + |   | II |
| 8#  | D | CCW | C | + | + | R+ | D | CCW | C | + | - | R+ |
| 9#  | D | CW | C | + | - | L+ | D |    | C | + |   | II |
| 10# | D | CW | C | + | - | L+ | D |    | C | + |   | II |
| 11# | D | CW | C | + | - | L+ | D |    | C | + |   | II |
| 12# | D | CCW | C | + | + | R+ | C |    | D | + |   | II |
| 13# | D | CCW | C | + | + | R+ | D | CCW | C | + | - | R+ |

**Table 1 Comparison of spin configuration of skyrmion bubbles between x-tilting and y-tilting.** D: divergent, C: convergent, CW: clockwise, CCW: counter-clockwise; L+: left-handed, type I bubble with positive polarity; R+: right-handed, type I bubble with positive polarity; L-: left-handed, type I bubble with negative polarity; R-: right-handed, type I bubble with negative polarity; II: Type-II bubble.

The magnetic features of marked bubbles in x-tilt series and y-tilt series in Figure 2a have been distinguished with above method and summarized in Table I. The corresponding 3D $B_x$ or $B_y$ component of each bubbles are shown in Supplementary information. The summary discloses that almost all of bubbles retain their positive polarity ($p+$) in x-tilt or y-tilt series, whether they vary the vorticity or not. We found that the transformation from type I bubbles to type II bubbles also obey this rule. It is well known that the field free skyrmion bubble is a metastable state and easily manipulates its spin chirality or topological texture under some stimuli, which is not only demonstrated here but also described in other literature[33]. In this work, a field cooling procedure is needed for the thin plate to generate the field free skyrmion bubbles. During field-cooling, the MnNiGa sample was magnetized by the applied field along the negative z direction, which results in the out-plane component inside individual bubble should be positive to survive in the MnNiGa matrix, otherwise it will merge into the matrix magnetic configuration and lost its special topological character. Thus, some external perturbation, such as rotating the specimen outside LTEM or the electron beam irradiation effect, may force the bubble to change its in-plane magnetic texture, but fail to upset the orientation of the out-plane component unless a strong external vertical field applied. As a result, the bubbles alter rather their vorticity than polarity to remain their existence in the MnNiGa matrix. It seems that rearranging the out-plane component of the magnetic moment of bubbles needs higher energy than reshaping the rotation direction of the in-plane components. Thus, a chirality oscillation of bubble, which transforms the swirling mode but maintains the polarity, may occur in centrosymmetric magnet Mn-Ni-Ga. This is a kind of pinning mechanism which holds the number of singular structures in the system.

## Discussion

Skyrmion bubbles (DDI skyrmions) have been one of the core components in commercial non-volatile memory in last century and lost their position with the development of hard disk techniques. However, the small size of individual bubble is still an attractive factor for the application in high density memory. With the emergence of the concept of skyrmions, topological property of the magnetic unit could be utilized for high density/speed memory. DMI skyrmions are believed to move easily under the DC current but the DDI bubbles prefer to reverse their vorticity because of the local pinning effect[13]. However, easily flipping the vorticity of bubbles without altering the polarity provides a new ability for a potential flexible operation in future memory, while the DMI skyrmion needs a vertical field to reverse its whirling axis in order to modify the spiral orientation.

In summary, the 3D magnetic configures of the skyrmion bubbles are revealed by the vector field tomography approach in LTEM. The reconstructed features confirm the surface state of bubbles. Both vorticity and polarity of the bubbles are discriminated and the type of the bubbles can be determined directly. We also illuminate the bubble type can be estimated from only one set of tilt data, simplifying the experimental procedure. The transformation of the bubbles is also clarified by the retrieved magnetic configure but the change is the chirality while the bubble maintains its polarity to keep the singular state in the matrix. The stable polarity of skyrmion bubble provides a potential application in future devices.

## Methods

***Sample fabrication.*** The thin plate for the LTEM characterization was cut from the bulk sample and prepared by mechanic polishing. The crystalline orientation of grain was checked by selected area electron diffraction (SAED), and only (001) facet was selected for further experiment. A field cooling (FC) manipulation was performed before the LTEM characterization to obtain field free magnetic bubbles aligned in [001] direction. The thickness of observing region is about 200 nm, determined from the reconstructed data.

***LTEM characterization.*** The skyrmion bubbles was studied by using JEOL-dedicated Lorentz TEM (JEOL2100F) equipped with high tilted holder and JEOL Cs-corrected ARM200F with Lorentz mode. No magnetic field was applied when performing the experiment. The coordinate is the laboratory frame where the direction of the electron beam is defined as z-axis and the corresponding perpendicular plane is x-y plane. The out-plane component of the bubble, or the easy [001]-axis of MnNiGa, was aligned as close as electron beam at initial position (tilt angle $\theta = 0$). The specimen was tilted around x-axis first in LTEM, then rotated 90 ° in plane manually outside the electron microscope and tilted around y-axis in LTEM again. Three images, over-focus, in-focus and under-focus, were acquired for each $\theta$. The tilt range was from -54.8 ° to 52.1 ° around x-axis and from -57.2 ° to 56.2 ° around y-axis, with the interval between 1-3 ° in order to avoid the image deterioration from strong diffraction contrast of the crystalline and ensure the correct reconstruction. The thickness of the characterized region was about 200 nm, estimated from the reconstructed 3D structure.

***3D structure reconstruction.*** The acquired images for each tilting angle θ including under, in and over focused images were processed by TIE approach to obtain phase image corresponding to θ, following the registration which remove the distortion among these three images because of the different focus condition. In TIE approach, the phase image can be calculated by the differential of images acquired at different focus conditions,

$$\varphi(x,y,z_0) = -k\nabla_{x,y}^{-2}\left\{\nabla_{x,y}\left[\frac{\nabla_{x,y}\nabla_{x,y}^{-2}\left(\frac{\partial I(x,y,z)}{\partial z}\right)}{I(x,y,z_0)}\right]\right\} \quad (1)$$

With the help of Fourier transform, the inverse Laplacian operator in eq. 1 can be presented feasibly as $\nabla_{x,y} \to \mathcal{F}^{-1}\left\{\frac{\mathcal{F}[\cdot]}{|q|^2}\right\}$, where q is the vector in x-y plane of frequency domain. In order to avoid the singularity when q = 0, a regularity parameter $q_0$ is needed $\nabla_{x,y} \to \mathcal{F}^{-1}\left\{\frac{\mathcal{F}[\cdot]|q|^2}{(|q|^2+q_0^2)^2}\right\}$. In this work, $q_0$ was 10 pixels to reach a compromise between depressing the low frequency noise efficiently and eliminating the artificial signal[14,15]. The generated phase image stack was aligned by the "Image Alignment" plug-in function in Gatan Microscopy Suite (GMS) automatically with a bandpass filter. A preferable method is to select a region enclosing the bubbles with stronger contrast as the reference to improve the alignment quality. The well aligned phase images were superimposed into one image and its Fourier transform (FT) provided a good indication to the tilting axis - the extension direction of diffused background[34]. Thereby the defined tilting axis inclined 62.7 ° left away from the vertical direction of raw image. Then the aligned phase image was rotated 62.7 ° clockwise to adjust the tilt axis along vertical direction of image since that direction is the default tilting axis in reconstruction algorithm. Before reconstruction, a horizontal differential with step of 10 pixels was made on the adjusted phase images to produce the component of magnetic induce along the tilting axis for each θ. Larger step size for differential is an effective concomitant filter to suppress the random noise. The 3D $B_x$ and $B_y$ matrix for different tilting axes were retrieved by the W-SIRT[35] program working in GMS, respectively.

For $B_z$ calculation, $B_x$ and $B_y$ matrixes were registered manually and reshaped to same size. From $\nabla \cdot \vec{B}(\vec{r}) = 0$, it is easy to get $\vec{k} \cdot \vec{B}(\vec{k}) = 0$ by the Fourier transform, where $\vec{B}(\vec{k})$ is the Fourier coefficient of $\vec{B}(\vec{r})$ and $\vec{k}$ is the grid in reciprocal space. Thus $B_z$ can be computed feasibly in frequency domain and transformed back to real space.

$$k_x B_x(\vec{k}) + k_y B_y(\vec{k}) + k_z B_z(\vec{k}) = 0 \quad (2)$$

$$B_z(\vec{k}) = -\frac{k_x B_x(\vec{k}) + k_y B_y(\vec{k})}{k_z} \tag{3}$$

$$B_z(\vec{r}) = \mathcal{F}^{-1}[B_z(\vec{k})] \tag{4}$$

It should be noted that for zero frequency in Eq. 3 the direct current component should be modified manually to satisfy the physical boundary conditions, for example $B_z$ in far region should be zero. The post-processing and visualization of the data were accomplished via Avizo®. Instead of magnetic inductions, sometimes magnetic moments are a proper physical quantity to outline the magnetic features. For brevity, these two concepts are used indiscriminately in following content since they are in same orientation within the specimen except the magnitudes differing by a material permeability.

## Acknowledgements

The authors thank Prof. Ying Zhang for the warm support in TEM characterization. This work was supported by the National Natural Science Foundation of China (Grant No. 11874410), the National Key R&D Program of China (2017YFA0303202, 2017YFA0206200) and the Strategic Priority Research Program of the Chinese Academy of Sciences (No. XDB33030200).

## Author Contributions

Y.Y. and W.H.W. supervised the project; Y.Y. and D.B. proposed the idea and designed the experiments; The MnNiGa polycrystal was synthesized by H.L; B.D. and J.J.L. prepared the lamella, performed the Lorentz-TEM measurement; Y.Y. analyzed the experimental data and completed reconstruction works; Y.Y. and D.B. wrote the manuscript after discussing data with other authors.

## Additional Information

**Supporting Information** is available at XXX

**Competing interests**: The authors declare no competing interests.

## Data availability

All the data that support the findings of this study are present in the paper and are available from the corresponding author upon reasonable request.

# Supplementary Information for

# The dynamic chirality flips of Skyrmion bubbles characterized by electron tomography

Yuan Yao[1,4]*, Bei Ding[1,4], Jinjing Liang[1,3], Hang Li[1], Xi Shen[1], Richeng Yu[1], Wenhong Wang[1,2,3]*

[1]Beijing National Laboratory for Condensed Matter Physics, Institute of Physics, Chinese Academy of Sciences, Beijing 100190, China

[2] Songshan Lake Materials Laboratory, Dongguan, Guangdong 523808, China

[3]University of Chinese Academy of Sciences, Beijing 100049, China

[4]These authors contributed equally: Yuan Yao and Bei Ding

**Corresponding Author**

*E-mail: yaoyuan@iphy.ac.cn
      wenhong.wang@iphy.ac.cn

**Supplementary Note 1. Raw images of tilting series**

The raw images of x-tilt and y-tilt series have been acquired in JEOL 2100F and the defocus value was 300 μm for the over-focus and under-focus images. The exposure time was fixed to 1 s for all images to ensure the uniform beam intensity.

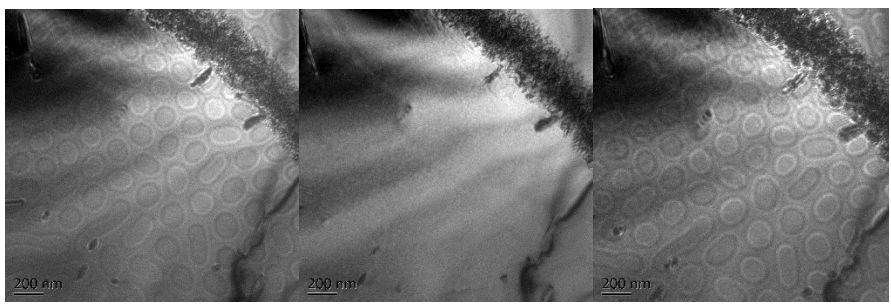

**Supplementary Figure 1** Under focus, in focus and over focus images of x-tilt 0.4

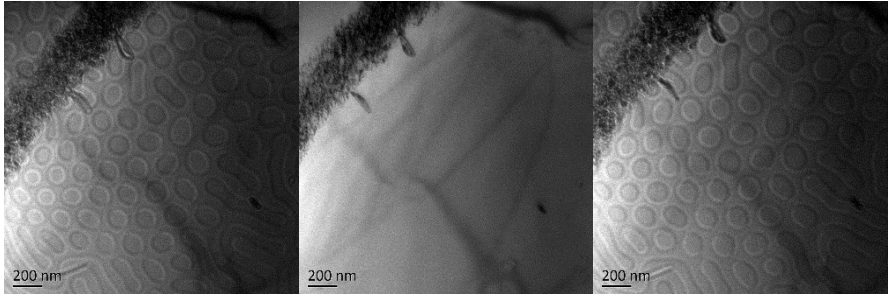

**Supplementary Figure 2** Under focus, in focus and over focus images of y-tilt 1.2

**Supplementary Note 2. 3D magnetic structure of typical type-I and type-II bubbles**

The micromagnetic simulations were carried out by using 3D object oriented micromagnetic framework (OOMMF) with the parameter used in previous report. [1] Magnetization matrix obtained from the OOMMF simulation was input into the home-made program to compute the exit wave of electron beam penetrating the simulated magnetic structures and the LTEM images were synthesized by the electron optical parameters, such as wavelength, defocus, energy dispersion envelope, etc.

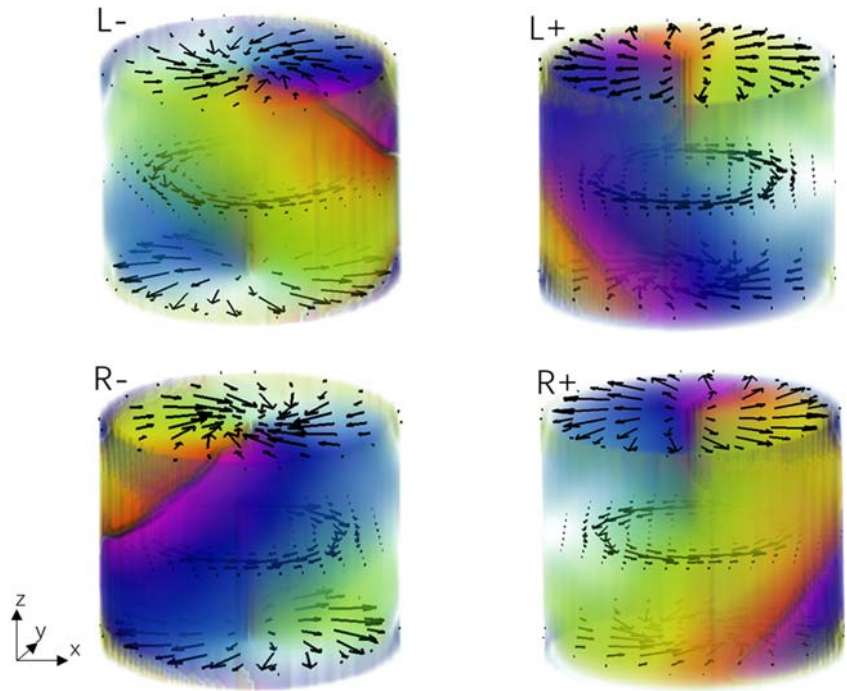

**Supplementary Figure 3** The rendered 3D structure of in-plane components of various Type-I bubbles.

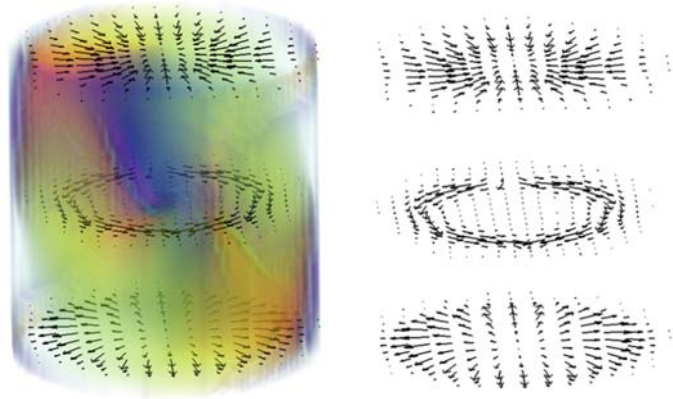

**Supplementary Figure 4** The rendered 3D structure of in-plane components of Type II bubbles.

**Supplementary Note 3. The type of bubbles recognized from the 3D $B_x$ or $B_y$ component**

The spin configurations of bubbles marked in Fig.2 are shown in this section.

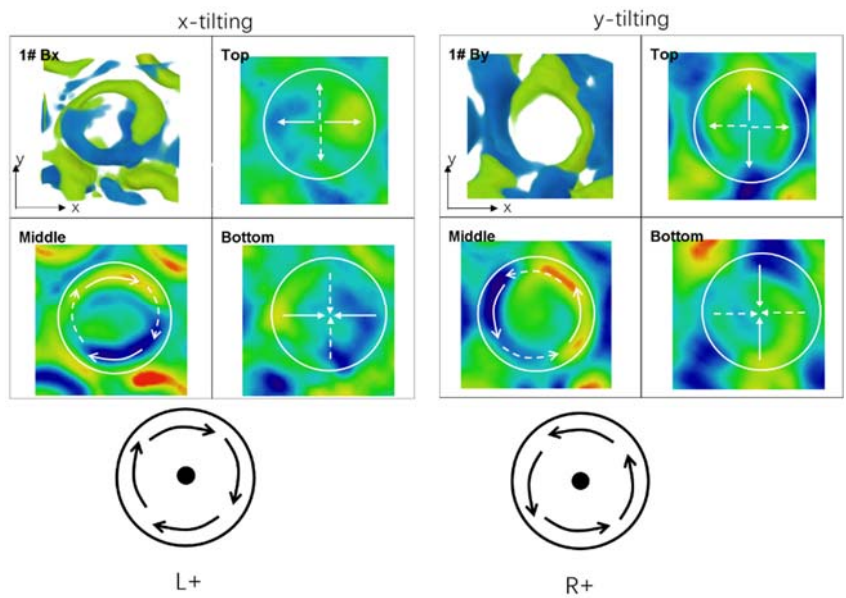

a

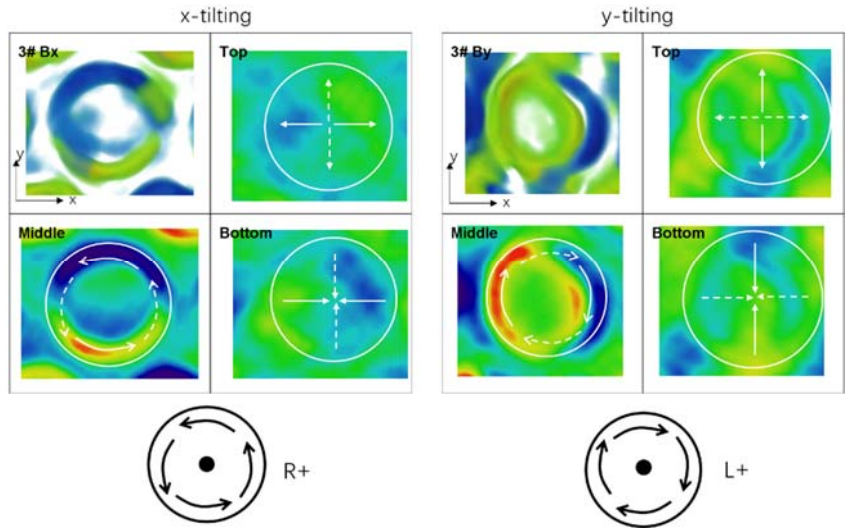

b

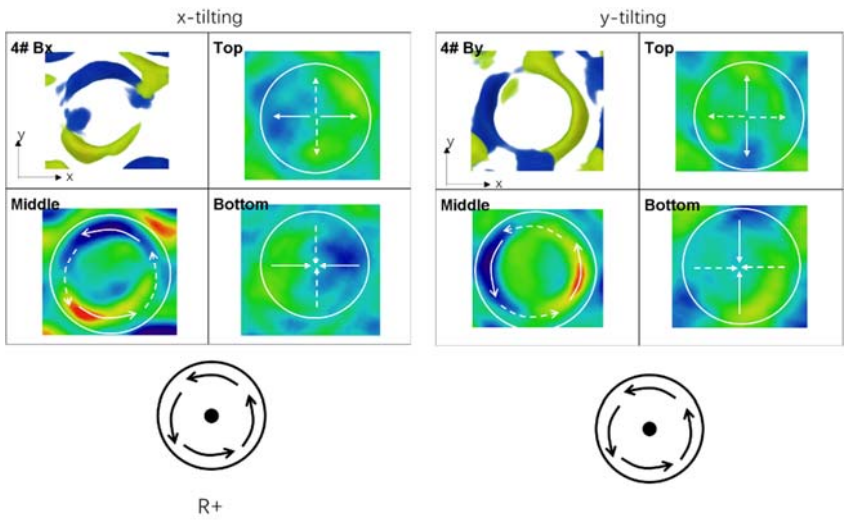

c

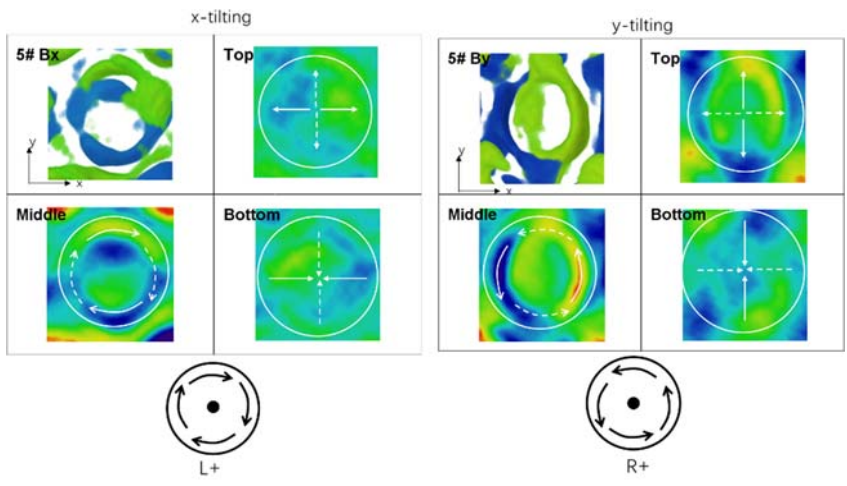

d

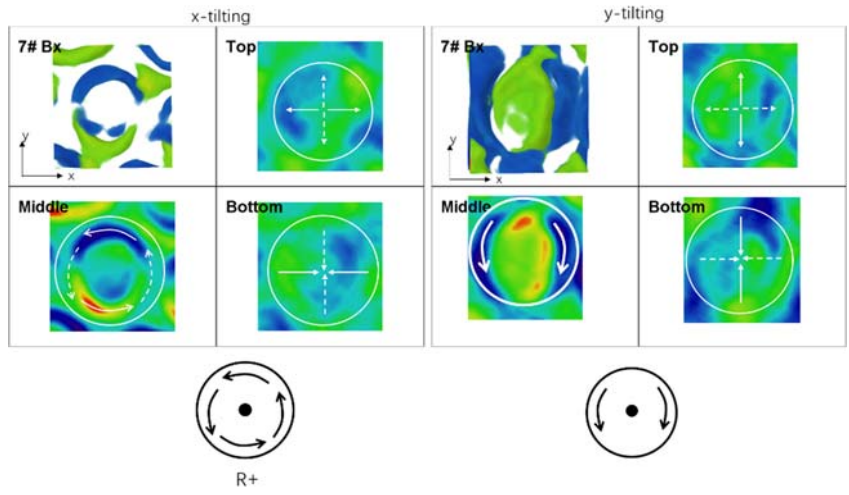

e

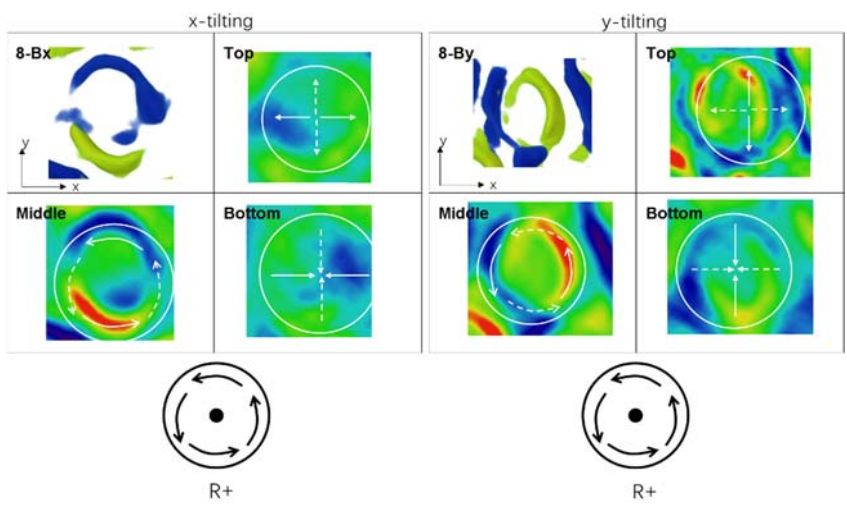

f

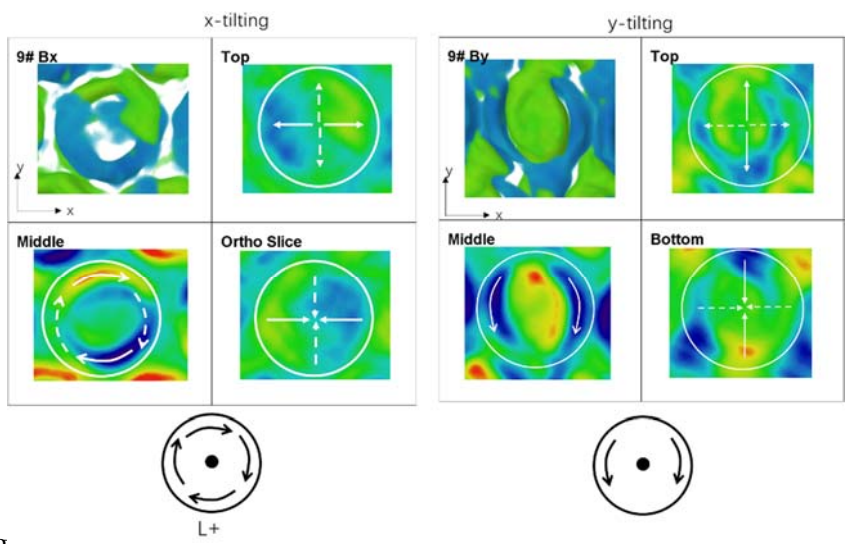

g

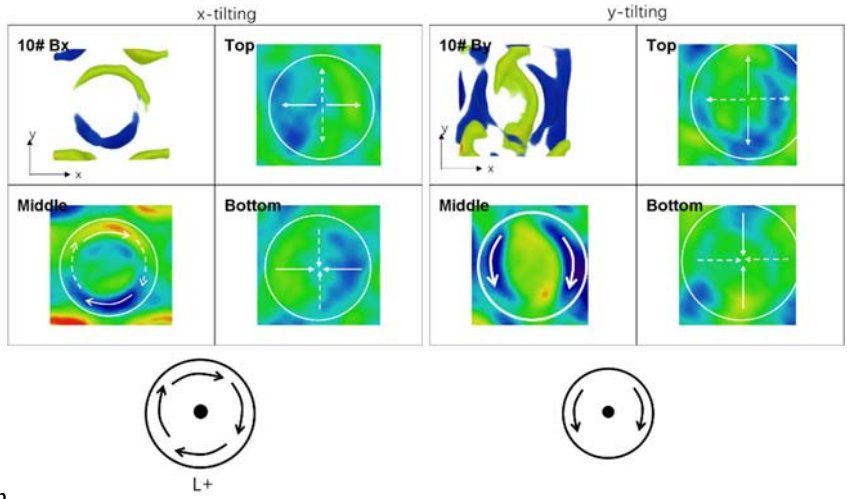

h

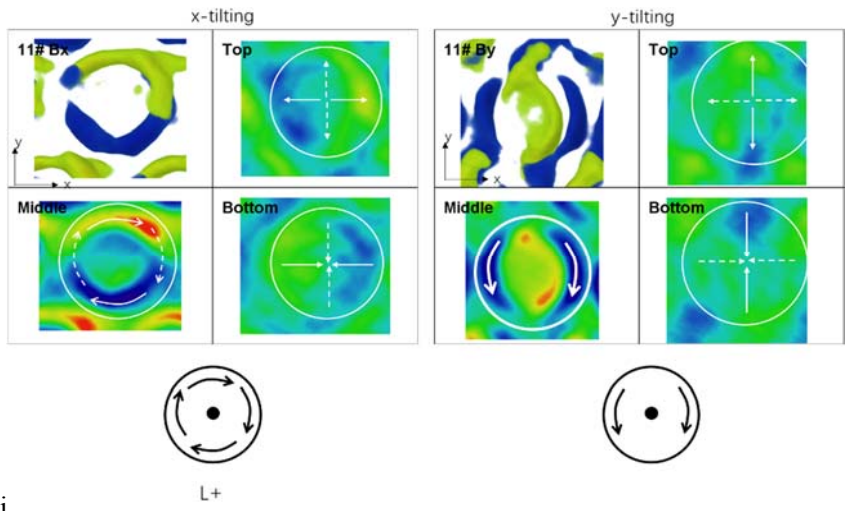

i

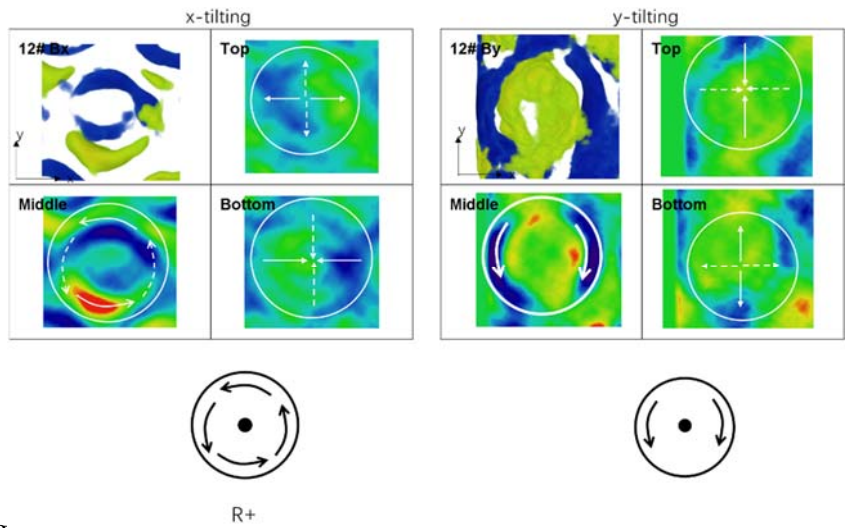

g

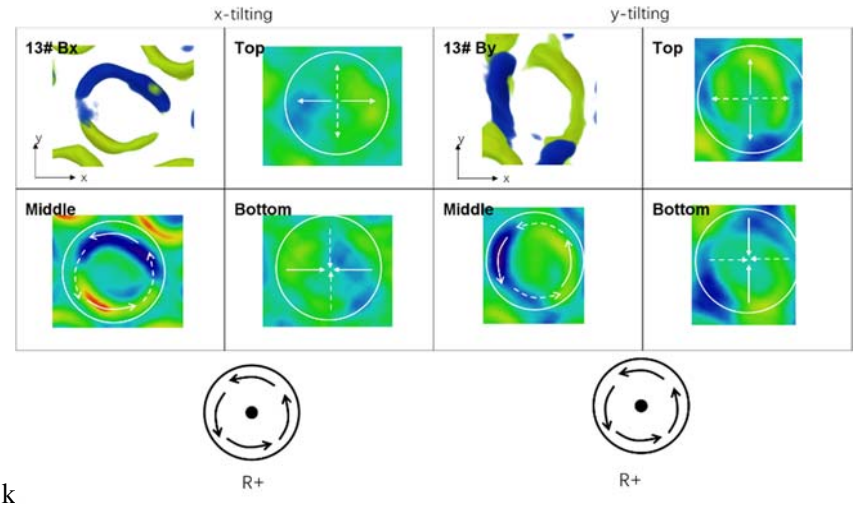

k

**Supplementary Figure 5 The type of bubbles recognized from the 3D $B_x$ or $B_y$ component**

**Moview 1 Bx of the bubbles**

**Moview 2 By of the bubbles**

**Moview 3 Bx and By of 2# bubble**

**Moview 4 Orientation mapping of in-plane component of 2# bubble and R+ model**

**Moview 5 Bx and By of 4# bubble**